\documentclass[12pt]{article}
\usepackage{color, pslatex}
\usepackage{amsmath}
\usepackage{amssymb}

\catcode`\@=11
\@addtoreset{equation}{section}

\begin{document}
\def\a{\alpha}
\def\b{\beta}
\def\c{\varepsilon}
\def\d{\delta}
\def\e{\epsilon}
\def\f{\phi}
\def\g{\gamma}
\def\h{\theta}
\def\k{\kappa}
\def\l{\lambda}
\def\m{\mu}
\def\n{\nu}
\def\p{\psi}
\def\q{\partial}
\def\r{\rho}
\def\s{\sigma}
\def\t{\tau}
\def\u{\upsilon}
\def\v{\varphi}
\def\w{\omega}
\def\x{\xi}
\def\y{\eta}
\def\z{\zeta}
\def\D{{\mit \Delta}}
\def\G{\Gamma}
\def\H{\Theta}
\def\L{\Lambda}
\def\F{\Phi}
\def\P{\Psi}
\def\S{\Sigma}

\def\rmd{{\rm d}}
\def\o{\over}
\def\beq{\begin{eqnarray}}
\def\eeq{\end{eqnarray}}
\newcommand{\gsim}{ \mathop{}_{\textstyle \sim}^{\textstyle >} }
\newcommand{\lsim}{ \mathop{}_{\textstyle \sim}^{\textstyle <} }
\newcommand{\vev}[1]{ \left\langle {#1} \right\rangle }
\newcommand{\bra}[1]{ \langle {#1} | }
\newcommand{\ket}[1]{ | {#1} \rangle }
\newcommand{\EV}{ {\rm eV} }
\newcommand{\KEV}{ {\rm keV} }
\newcommand{\MEV}{ {\rm MeV} }
\newcommand{\GEV}{ {\rm GeV} }
\newcommand{\TEV}{ {\rm TeV} }
\def\slash#1{\ooalign{\hfil/\hfil\crcr$#1$}}
\def\diag{\mathop{\rm diag}\nolimits}
\def\Spin{\mathop{\rm Spin}}
\def\SO{\mathop{\rm SO}}
\def\O{\mathop{\rm O}}
\def\SU{\mathop{\rm SU}}
\def\U{\mathop{\rm U}}
\def\Sp{\mathop{\rm Sp}}
\def\SL{\mathop{\rm SL}}
\def\tr{\mathop{\rm tr}}

\newcommand\red[1]{{\textcolor{red}{#1}}}

\begin{titlepage}
\vskip-1.0cm
\begin{flushright}
{\tt ArXiV:0905.2940}\\
SNUST 090501\\
UCB-PTH-09/16
\end{flushright}
\vskip0.25cm
\centerline{\Large \bf  Observables and Correlators}
\vspace{0.25cm}
\centerline{\Large \bf in}
\vspace{0.25cm}
\centerline{\Large \bf Non-Relativistic ABJM Theory}
\vspace{1.cm}
\centerline{\large
Yu Nakayama $^a$, \,\,\,  Soo-Jong Rey $^{b,c}$}
\vspace{1.cm}
\centerline{\sl $^a$ Berkeley Center for Theoretical Physics \& Department of Physics}
\vskip0.15cm
\centerline{\sl University of California, Berkeley CA 94720-7300 {\rm USA}}
\vskip0.25cm
\centerline{\sl $^b$ School of Physics and Astronomy \& Center for Theoretical Physics}
\vskip0.15cm
\centerline{\sl Seoul National University, Seoul 151-747 {\rm KOREA}}
\vskip0.25cm
\centerline{\sl $^c$ Kavli Institute for Theoretical Physics, Santa Barbara CA 93106 {\rm USA}}
\vskip0.25cm
\centerline{\tt
nakayama@berkeley.edu, \,\,\, sjrey@snu.ac.kr}
\vspace{0.75cm}
\centerline{ABSTRACT}
\vspace{0.75cm}
\noindent
Non-relativistic ABJM theory is defined by Galilean limit of mass-deformed ${\cal N}=6$ Chern-Simons theory. Holographic string theory dual to the theory is not known yet. To understand features candidate gravity dual might exhibit, we examine local and nonlocal physical observables and their correlations in the non-relativistic ABJM theory. We show that gauge invariant local observables correspond to zero-norm states and that correlation functions among them are trivial. We also show that a particular class of nonlocal observables, Wilson loops, are topological in the sense that their correlation functions coincide with those of pure Chern-Simons theory. We argue that the theory is nevertheless physical and illustrate several physical observables whose correlation functions are nontrivial. We also study quantum aspects. We show that Chern-Simons level is finitely renormalized and that dilatation operator is trivial at planar limit. These results all point to string scale geometry of gravity dual and to intriguing topological and tensionless nature of dual string or M theory defined on it.
\end{titlepage}

\setcounter{page}{1}

\section{Introduction}

Non-relativistic version of the anti-de Sitter (AdS) / conformal field theory (CFT) correspondence \cite{Son:2008ye}\cite{Balasubramanian:2008dm} is an interesting new horizon to string theory.
The correspondence offers a new arena of theoretical physics by providing a completely new approach to strongly correlated condensed matter systems such as cold atoms in optical trap, high $T_c$ superconductor and quantum Hall effects. Conversely, the correspondence opens up an exciting possibility that experimental as well as theoretical understanding of strongly correlated condensed matter systems may offer a new insight into quantization of gravity and string theory in a suitable dual background.

Identification of physical observables in both the AdS and the CFT sides is the starting point for quantitative study of the correspondence. In the relativistic case, a prescription to compute the correlation functions of local observables (so-called GKPW relation \cite{GKP}\cite{W}) and of nonlocal observables such as Wilson loops \cite{Rey:1998ik}\cite{Maldacena:1998im} were proposed immediately after the advent of the AdS/CFT correspondence \cite{Maldacena:1997re}.
Drawing an analogy to the relativistic AdS/CFT correspondence, a prescription to compute the correlation functions in the non-relativistic AdS/CFT correspondence was proposed recently \cite{Son:2008ye}\cite{Balasubramanian:2008dm}\cite{Fuertes:2009ex}. However, in all cases studied so far, 'boundary' of the spacetime dual to the non-relativistic CFT turned out singular, so it was not obvious how to set up a sensible prescription. In this regard, though partial success was reported, the proposed prescriptions are far from being rigorous and should be considered at best heuristic.

In the non-relativistic AdS/CFT correspondences studied so far, there is no concrete Lagrangian description of the dual conformal field theory. This is a serious drawback against testing the correspondence. One promising and distinguished candidate in this direction is the non-relativistic M2-brane theory. The construction is simple: we begin with the worldvolume gauge theory of $N$ multiply stacked relativistic M2-branes at an orbifold singularity \cite{Aharony:2008ug}, introduce a suitable supersymmetric mass deformation as in \cite{Hosomichi:2008jd} \cite{Gomis:2008vc}, and then take an appropriate non-relativistic limit. The resultant action, which we call as the non-relativistic ABJM theory, has been reported in \cite{Nakayama:2009cz}, and the (super)symmetry has been investigated.

In this paper, we study physical observables and their correlators in the non-relativistic ABJM theories with gauge group U($N) \times $U($N$) and Chern-Simons level $k$. The theory has 10 kinematical, 2 dynamical and 2 conformal supersymmetries, so these observables are classifiable according to their supermultiplet structures. We shall various surprising features that distinguishes the non-relativistic ABJM theory from the relativistic counterpart. We first show that {\sl all} local gauge invariant observables are zero-norm states and hence {\sl all} correlation functions among themselves are trivial. We will further show that a special class of non-local observables, the Wilson loops and the `t Hooft loops, are {\sl topological} in the sense that their correlation functions coincide with those of the pure Chern-Simons theory. By the holography, this suggests a peculiar topological nature of the string theory defined on gravity dual.

We emphasize that the non-relativistic ABJM theory is {\sl not} a topological field theory but a physical theory with intricate multi-particle dynamics. For example, the theory shows non-trivial S-matrices. The claim here is that if we restrict ourselves to correlation functions involving a class of simple (local or non-local) gauge invariant observables studied in the context of the relativistic AdS/CFT correspondence, they exhibit topological characteristic. Indeed, we find physically nontrivial correlation functions once we take account of less familiar observables involving monopole operators and multi-local operators. In discerning topological versus physically nontrivial observables, we point out eigenvalue of these operators to the mass operator in the underlying Schrodinger superalgebra plays a prominent role.

Related, we emphasize that quantum dynamics of the non-relativistic ABJM theory gives rise to {\sl finite} renormalization of the Chern-Simons level $k \rightarrow k + N$. We also emphasize that the dilatation operator of the theory acting on a single trace state is an identity to all orders in `t Hooft coupling perturbation theory. This implies that there is no dynamical excitation on the string worldsheet. From these, we learn a great deal of yet-to-be-found candidate string theory and gravity dual to the non-relativistic ABJM theory. In particular, we see that the gravity dual must have a geometry of curvature at string scale and string theory defined on it must exhibit tensionless and topological characteristic.

We organized this paper as follows. In section 2, we review the non-relativistic ABJM theory and its supersymmetries. In section 3, utilizing the established operator-state correspondence to this theory, we study properties of local gauge invariant observables. We show that all local operators have zero norm and their correlation functions vanish. We also discuss gravity dual of this property and classify supersymmetry protected local operators. In section 4, extending to nonlocal observables, we study Wilson loops. We show that they are topological: their correlation functions are exactly the same as the pure Chern-Simons theory. From this, we also find that string dual to the Wilson loop must be tensionless. We also classify supersymmetric Wilson loops, which are natural extensions of the chiral primary local operators. In section 5, we argue that there are also observables of nonzero norm. We also study correlation functions involving these observables and show that they are nontrivial -- neither zero nor topological. In section 6, we highlight several quantum aspects of the theory. We argue for the shift of the Chern-Simons coefficient and discuss its implications. We also point out topological nature of non-relativistic spin chain on which dilatation operator acts as an identity operator. This implies that there is no excitation on the worldsheet of string dual. In section 7, we summarize our results and discuss various implications and points for further study.

\section{Non-relativistic ABJM Theory}
The non-relativistic ABJM theory with fourteen supercharges was first derived in \cite{Nakayama:2009cz} by taking non-relativistic limit of the mass deformed ABJM theory \footnote{In \cite{Nakayama:2009cz}, less supersymmetric ABJM theories were also proposed. In this paper, we focus on the maximally supersymmetric case.} (see also \cite{Lee:2009mm}). Subsequently, a superfield formulation with manifest dynamical supersymmetry (SUSY) was developed in \cite{Nakayama:2009ku}. In this paper, we follow the conventions and the notations used in \cite{Nakayama:2009cz}.

\subsection{action}
The Chern-Simons part of the non-relativistic ABJM theory is the same as the relativistic theory and is given by a pair of U$(N)$ Chern-Simons action with levels $k$ and $-k$, respectively:
\begin{eqnarray}
S_{\rm CS} = \frac{k}{4\pi} \int\! \rmd t \rmd^2 x \,\, \epsilon^{mnp}\,
\mathrm{Tr} \left[ A_m \partial_n A_p
+ \frac{2i}{3}A_m A_n A_p
 - \overline{A}_m \partial_n \overline{A}_p
-\frac{2i}{3} \overline{A}_m \overline{A}_n \overline{A}_p \right] \ . \label{CSa}
\end{eqnarray}
Here, $A_m, \overline{A}_m$ are gauge potentials of U$(N)\times$U($N$). Gauge invariance of quantum dynamics requires $k$ integer-valued. Matter fields consist of bosons $\phi^A$ and fermions $\psi_A$, transforming as $({\bf N}, \overline{\bf N})$ under the gauge group U$(N)\times$U($N$). Here, $A = (a,a')$ denotes the global SU(2)$\times$ SU(2) indices: $(1,2,1',2')$. Note that in mass deformed theory the parity conjugation exchanges not only the two U(N) gauge groups but also the two SU(2) R-symmetry groups.

Mass deformation of the ABJM theory preserving the ${\cal N}=6$ Poincar\'e supersymmetry (but breaking Osp$(6 \vert 4)$ superconformal invariance) introduces a mass scale $m$. The deformation breaks the SU(4) R-symmetry to SU(2)$\times$SU(2)$\times$U(1). The U(1) is generated by the generator diag.$(+{1 \over 2},+{1 \over 2},-{1 \over 2},-{1 \over 2})$. From the viewpoint of ${\cal N}=2$ superspace formulation, the mass deformation of the relativistic ABJM theory arises only from the D-term contribution to the scalar potential, none from the F-term. This may be seen as follows. Assume that it arises from the F-term $[W(\phi^a, \phi^{a'})]_{\theta^2}+$(h.c.). As the mass deformation preserves SU(2)$\times$SU(2) subgroup, the unique choice of holomorphic operators of scaling dimension-2 are $m \mbox{Tr} (\phi^a \epsilon_{ab} \phi^b \pm \phi^{a'} \epsilon_{a'b'} \phi^{b'})$. However, these operators can not be present since they violate U(1) invariance and moreover vanishes identically upon taking trace over the gauge group.

The non-relativistic limit is then obtained by taking the limit $c \rightarrow \infty$, and keep in each fields the particle mode only --- it turned out this is required to retain maximal supersymmetry. Denoting the mass as $m$, action for the non-relativistic kinetic term is given by\footnote{Our notation for spacetime and spatial coordinates are $\mu = 0, 1, 2$ and $i=1,2$, respectively.
We introduce $V_\pm = V^\pm V_1 = \pm iV_2$ for every vector fields. In particular, $D_\pm = D_1 \pm iD_2$, where ${\bf D} = (D_1, D_2)$ are spatial components of the gauge covariant derivative of U$(N)\times$U($N$). Here and throughout, we also rescaled the matter fields appropriately and put the coupling constant ${k / 4 \pi}$ as overall factor to the action.}
\begin{align}
S_{\rm kin} &= {k \over 4 \pi} \int\! \rmd t \rmd^2x\,\, \left[\mathrm{Tr}
(\phi^\dagger_A\,  i D_0 \phi^A)
- \frac{1}{2m} \mathrm{Tr} ({\bf D} \phi_A)^\dagger ({\bf D} \phi^A) \right. \cr
& \quad \qquad + \left. \mathrm{Tr} (\psi^{\dagger A} \,\, i D_0 \psi_A)
+ \frac{1}{2m} \mathrm{Tr} (\psi^{\dagger a} D_- D_+ \psi_a
+ \psi^{\dagger a'} D_+ D_- \psi_{a'}) \right] . \label{kine}
\end{align}
The last term is the Pauli interaction term, and can be rewritten as
\begin{align}
&  \frac{1}{2m} \mathrm{Tr} (\psi^{\dagger a} D_- D_+ \psi_a
+ \psi^{\dagger a'} D_+ D_- \psi_{a'}) \cr
=& \frac{1}{2m} \mathrm{Tr}\left[
\psi^{\dagger a} {
\left({\bf D}^2\psi_a - F_{12}\psi_a + \psi_a\overline{F}_{12} \right)
}
\right]
+ \frac{1}{2m} \mathrm{Tr}\left[\psi^{\dagger a'} {
\left({\bf D}^2 \psi_{a'}  + F_{12}\psi_{a'}  -\psi_{a'}  \overline{F}_{12} \right)
}
\right] \ . \notag
\end{align}
The action for the quartic scalar self-interactions is given by \footnote{A convenient convention would be
to put $2 m = 1$. This will eliminate factors of $2m$ in the action and supersymmetry transformations.
For a book-keeping convenience, however, we shall keep this factor explicit hereafter.}
\begin{eqnarray}
S_{\rm bose} = \frac{k}{4 \pi} \int\! \rmd t \rmd^2x\,\,
{1 \over 4 m} \, \mathrm{Tr}
\left(\phi^a\phi^\dagger_{[a} \phi^b\phi^\dagger_{b]}
- \phi^{a'}\phi^\dagger_{[a'} \phi^{b'} \phi_{b']}^\dagger\right) \ ,
\label{bosa}
\end{eqnarray}
whereas the action for quartic scalar-fermion interactions is given by
\begin{align}
S_{\rm fermi} = \frac{k}{4\pi }&\int\! \rmd t \rmd^2 x\,  {1 \over 4m} \, \mathrm{Tr} \left[
(\phi^\dagger_a\phi^a + \phi^\dagger_{a'} \phi^{a'})(\psi^{\dagger
b}\psi_b - \psi^{\dagger b'} \psi_{b'})
\right.
\nonumber \\
& \hskip2.3cm +(\phi^a\phi^\dagger_a
+\phi^{a'}\phi^\dagger_{a'} )(\psi_b \psi^{\dagger b} - \psi_{b'}
\psi^{\dagger b'} )  \nonumber \\
&  \hskip2.3cm - 2 \phi^a\phi^\dagger_b
\psi_a\psi^{\dagger b} + 2 \phi^{a'} \phi^\dagger_{b'} \psi_{a'}
\psi^{\dagger b'}
- 2 \phi^\dagger_a \phi^b\psi^{\dagger a}\psi_b +2
\phi^\dagger_{a'} \phi^{b'} \psi^{\dagger a'} \psi_{b'} \nonumber \\
& \hskip2.3cm
-i\epsilon^{ab}\epsilon^{c'd'} \phi^\dagger_a \psi_b \phi^\dagger_{c'}
\psi_{d'}
- i \epsilon^{bc}\epsilon^{a'd'}\phi^\dagger_{a'} \psi_b
\phi^\dagger_c\psi_{d'} \nonumber \\
& \hskip2.3cm
+i\epsilon^{a'b'}\epsilon^{cd} \phi^\dagger_{a'}
\psi_{b'} \phi^\dagger_{c} \psi_{d}+
i\epsilon^{b'c'}\epsilon^{ad}\phi^\dagger_{a}
\psi_{b'}\phi^\dagger_{c'}\psi_{d} \nonumber \\
&
\hskip2.3cm
+i\epsilon_{ab}\epsilon_{c'd'} \phi^a \psi^{\dagger b} \phi^{c'}
\psi^{\dagger d'} + i \epsilon_{bc}\epsilon_{a'd'}\phi^{a'}
\psi^{\dagger b} \phi^c\psi^{\dagger d'} \nonumber \\
& \hskip2.3cm \left. -i\epsilon_{a'b'}\epsilon_{cd}
\phi^{a'} \psi^{\dagger b'} \phi^{c} \psi^{\dagger d}-
i\epsilon_{b'c'}\epsilon_{ad}\phi^{a} \psi^{\dagger
b'}\phi^{c'}\psi^{\dagger d} \right] \ . \label{fera}
\end{align}
The non-relativistic ABJM theory is then defined
by the combined action of \eqref{CSa}, \eqref{kine}, \eqref{bosa} and \eqref{fera}.
Notice that the theory does not conserve fermion number. Therefore, despite being non-relativistic, the theory permits processes in which a pair of fermions transmutes to a pair of bosons and vice versa.

\subsection{kinematical, dynamical and conformal supersymmetries}
We now analyze supersymmetries of the non-relativistic ABJM theory. In general, the supersymmetry in non-relativistic system is decomposed into the leading component in the non-relativistic limit, called kinematical supersymmetry, and the remaining component, called dynamical supersymmetry. The action is invariant under the following ${\cal N}=6$ kinematical supersymmetry transformations:
\begin{align}
& \delta \phi^{A} = \sqrt{2m} \left(\xi^{Ab} \psi_b + i \hat{\xi}^{Ab'} \psi_{b'} \right)\cr
& \delta \phi^\dagger_{A} = \sqrt{2m} \left(-\hat{\xi}_{Ab} \psi^{\dagger b} + i\xi_{Ab'} \psi^{\dagger b'} \right) \cr
& \delta{\psi}_{a} = \sqrt{2m} \left(\hat{\xi}_{aB} \phi^B\right) \cr
& \delta\psi_{a'} = \sqrt{2m} \left(-i\xi_{a'B}\phi^B \right) \cr
& \delta{\psi}^{\dagger a} = \sqrt{2m} \left(\xi^{aB}\phi^\dagger_B \right) \cr
& \delta{\psi}^{\dagger a'} = \sqrt{2m} \left( i\hat{\xi}^{a' B} \phi^\dagger_B\right) \cr
& \delta{A_0} = +  \sqrt{2m} \frac{1}{4m} \left(\phi^A (\hat{\xi}_{Ab} \psi^{\dagger b} + i \xi_{A b'}  \psi^{\dagger b'}) + (\xi^{aB} \psi_{a} - i\hat{\xi}^{a' B} \psi_{a'})\phi_{B}^\dagger \right)  \cr
& \delta{\overline{A}}_0 = -\sqrt{2m} \frac{1}{4m} \left( (\hat{\xi}_{aB} \psi^{\dagger a} + i \xi_{a' B}\psi^{\dagger a'}) \phi^B + \phi_A^\dagger (\xi^{Ab} \psi_b - i\hat{\xi}^{A b'} \psi_{b'} ) \right) \ \cr
& \delta{A}_i = 0 \cr
& \delta\overline{A}_i = 0 \ .
\label{kins}
\end{align}
Here, $\xi^{AB}$ denotes 6 complex spinor parameters of the kinematical supersymmetry. It turns out that only 5 of them are independent. To see this, we first denote them as $\varepsilon_i$ ($i=1, \cdots, 6$):
\begin{eqnarray}
\xi^{AB} = (\Gamma^{*i})^{AB}\epsilon^i , \quad \hat{\xi}^{AB} = (\Gamma^{*i})^{AB} \epsilon^{*i} , \quad \xi_{AB} = (\Gamma^i)_{AB} \epsilon^i , \quad \hat{\xi}_{AB} = (\Gamma^i)_{AB} \epsilon^{*i} \ .
\end{eqnarray}
the gamma matrices $\Gamma^i_{AB}$ are taken as
\begin{eqnarray}
& \ \ \Gamma^1 = \ \sigma_2 \otimes \mathbb{I}\, , \qquad \ \ \ \Gamma^2 = -i\sigma_2 \otimes \sigma_3\, , \nonumber \\
& \ \Gamma^3 = i\sigma_2 \otimes \sigma_1\, , \qquad \Gamma^4 = -\sigma_1 \otimes \sigma_2\, ,  \nonumber \\
& \! \Gamma^5 = \sigma_3 \otimes \sigma_2\, , \qquad \ \Gamma^6 = -i \ \mathbb{I} \otimes \sigma_2 \, .
\end{eqnarray}
These chiral SO(6) gamma matrices are the intertwiner between the
SU(4) antisymmetric representation (with the reality condition) and the
SO(6) (real) vector representation. Note that $\frac{1}{2}\epsilon^{ABCD} \Gamma_{CD}^i = -(\Gamma^{i*})^{AB}$.
From these, one finds in the large component of $m \rightarrow \infty$ that the spinor parameters $\varepsilon_1$ and $\epsilon_2$ are not mutually independent since $\epsilon_1 - i \epsilon_2$ component is sub-leading as $m \rightarrow \infty$, so there are only 5 independent kinematical supersymmetry transformations. We denote the corresponding supercharges as $Q_0$ and $Q_m$ ($m=1,\cdots,4$). Under the SU(2)$\times$SU(2) R-symmetry, they transform as $({\bf 1}, {\bf 1})$ and $({\bf 2}, {\bf 2})$, respectively.

The combination $(\epsilon_1 - i\epsilon_2)$ of the spinor parameters still yield a residual contribution to the smaller component. It generates the dynamical supersymmetry transformations:
\begin{align}
& \delta \phi^{A} = \frac{1}{\sqrt{2m}} \left(-i\hat{\zeta}^{Ab}D_+\psi_b + \zeta^{Ab'} D_-\psi_{b'} \right)\cr
& \delta \phi^\dagger_{A} = \frac{1}{\sqrt{2m}} \left(-i{\zeta}_{Ab} D_-\psi^{\dagger b} -\hat{\zeta}_{Ab'} D_+\psi^{\dagger b'} \right) \cr
& \delta{\psi}_{a} = \frac{1}{\sqrt{2m}} \left(-i{\zeta}_{aB}  D_-\phi^B\right) \cr
& \delta\psi_{a'} = \frac{1}{\sqrt{2m}} \left(-\hat{\zeta}_{a'B}D_-\phi^B \right) \cr
& \delta{\psi}^{\dagger a} = \frac{1}{\sqrt{2m}} \left(i\hat{\zeta}^{aB}D_+\phi^\dagger_B \right) \cr
& \delta{\psi}^{\dagger a'} = \frac{1}{\sqrt{2m}} \left( -{\zeta}^{a' B} D_-\phi^\dagger_B\right) \cr
& \delta{A_0} = + \frac{1}{\sqrt{2m}} \frac{1}{4m} \left(\phi^A (-i{\zeta}_{Ab} D_-\psi^{\dagger b} + \hat{\zeta}_{A b'} D_+\psi^{\dagger b'}) + (i\hat{\zeta}^{aB} D_+ \psi_{a} + {\zeta}^{a' B} D_- \psi_{a'})\phi_{B}^\dagger \right)  \cr
& \delta A_+ = + \frac{1}{\sqrt{2m}} \left( \phi^A \zeta_{Ab} \psi^{\dagger b} - i \zeta^{a' B} \psi_{a'} \phi^\dagger_{B} \right) \cr
& \delta A_- = + \frac{1}{\sqrt{2m}} \left(\hat{\zeta}^{aB} \psi_a \phi^\dagger_B + i\phi^A \hat{\zeta}_{Ab'} \psi^{\dagger b'} \right)  \cr
& \delta{\overline{A}}_0 = -\frac{1}{\sqrt{2m}} \frac{1}{4 m} \left( (-i{\zeta}_{aB}D_- \psi^{\dagger a} +  \hat{\zeta}_{a' B}D_+\psi^{\dagger a'}) \phi^B + \phi_A^\dagger (i\hat{\zeta}^{Ab} D_+\psi_b + {\zeta}^{A b'} D_-\psi_{b'} ) \right) \ \cr
& \delta\overline{A}_+ = -\frac{1}{\sqrt{2m}} \left(\zeta_{aB} \psi^{\dagger a} \phi^{B} - i\phi_A^\dagger \zeta^{Ab'} \psi_{b'} \right) \cr
& \delta\overline{A}_- = -\frac{1}{\sqrt{2m}} \left(i\hat{\zeta}_{a'B} \psi^{\dagger a'} \phi^B + \phi^\dagger_A \hat{\zeta}^{Ab} \psi_b \right) \ . \label{dyns}
\end{align}
the spinor parameter of dynamical supersymmetry is $\zeta^{AB}$
\begin{eqnarray}
\zeta^{AB} = (\Gamma^{*i})^{AB}\epsilon^i , \ \hat{\zeta}^{AB} = (\Gamma^{*i})^{AB} \epsilon^{*i} , \ \zeta_{AB} = (\Gamma^i)_{AB} \epsilon^i , \ \hat{\zeta}_{AB} = (\Gamma^i)_{AB} \epsilon^{*i} \ .
\end{eqnarray}
with only $i=1$ and $i=2$ contributing to the transformations. We denote the dynamical supercharge as ${\cal Q}$. Under the SU(2)$\times$SU(2) R-symmetry, the supercharge ${\cal Q}$ is a singlet.

Finally, associated to the two dynamical supersymmetries, there are two conformal supersymmetries. In fact, they are imperative in order for kinematical and dynamical supersymmetries form non-relativistic supersymmetry algebra.

\subsection{non-relativistic superconformal algebra}

We now summarize the non-relativistic superconformal algebra with fourteen supercharges realized in the non-relativistic ABJM theory. The bosonic part is nothing but the Schr\"odinger algebra -- non-relativistic conformal algebra: \cite{Hagen:1972pd} - \cite{Hussin:1986cc}:
\begin{align}
& i[J,P_+] = -i P_+ \ , \ \ \ \ \ i[J,P_-] = + i P_- \ , \ \ \ \ \ \ i[J,G_+] = - iG_+ \ , \ \ \ \ i[\ J ,G_-] = + iG_- \ ,\cr
& i[H,G_+] = + P_+ \ ,  \ \, \ \ \ i[H,G_-] = + P_- \ , \ \ \ \ i[K,P_+] = - G_+ \ ,  \ \ \ \ \ i[\ K,P_-] = - G_- \ ,\cr
& i[D,P_+] = - P_+ \ , \ \ \ \ \ i[D,P_-] = - P_- \ , \ \ \ \ \ i[D,G_+] = + G_+ \ ,  \ \ \ \ \ i[\ D,G_-] = + G_- \ ,\cr
& i[H, D] = \ 2H \ , \ \ \ \ \ \ \ \ i[H,K] =  \ D \ , \ \ \ \ \ \ \ \ \ \ i[D,K] = \ 2K \ ,  \ \ \ \ \ \ \ \ \ \ i[P_+,G_-] = \ 2M \ .
\end{align}
In our notation, $H$ is the non-relativistic Hamiltonian, $P_\pm$ are the momentum, $J$ is the U(1) angular momentum, $D$ is the dilatation, $K$ is the special conformal transformation, and $G_z, G_{\overline{z}}$ are the Galilean boost generators. Moreover, $M$ is the total mass generator
\begin{eqnarray}
M = m \int \rmd^2 x \,\, \rho \qquad \mbox{where} \qquad \rho = \mbox{Tr} (\phi^\dagger_A \phi^A + \psi^{\dagger A} \psi_A)
\end{eqnarray}
and $\rho$ measures the particle number density of the non-relativistic matter fields. An important point for later discussions is that $\rho$ is also the matter part to the electric charge of overall U(1) in U$(N)\times$U($N$) gauge group.

The non-vanishing fermionic part of the non-relativistic superconformal algebra is
\begin{align}
& \{Q_0 , Q_{0}^* \} \ = \ 2  M \ , \ \ \  \{ \ Q_{m}, \ Q_{n}^* \} \ = \ 2 M \delta_{mn}+ i \, 2m R_{mn} \ , \cr
& \{ \ {\cal Q},  {\cal Q}^*\ \} \ = \ H \ , \ \ \  \ \ \{Q_0,{\cal Q}^*\} = P_- \ , \ \ \ \ \ \ \ \{ {\cal Q},{Q}^{0*}\} = P_+ \ , \ \cr
& \{ \ {\cal S}, \ {\cal S}^*\ \} \ = \ + K \ , \ \ \ \{ {\cal S},Q_{0}^* \} = -G_+ \ ,  \ \ \ \ \{ {\cal S},{\cal Q}^*\} = \frac{i}{2}(iD - J +\frac{3}{2}R) \ , \cr
& i [ \ J, \ Q_{0} \ ] \ =  \frac{i}{2}Q_0 \ , \ \
\ i [ \ J, \ Q_{m} \ ] \ =  \frac{i}{2}Q_m \ , \ \ \ i [ \ J, \ {\cal Q} \ ] \ = \ -\frac{i}{2} {\cal Q} \ , \cr
& i [  G_-, {\cal Q} \ ] \ =  - Q_{0} \ ,
\ \ \ i  [ \ D, \ {\cal Q} \ ]  =   - {\cal Q} \ ,
\ \ \ \ \ \ i [ \ K,\ {\cal Q} \ ] \ = \ + \ {\cal S} \ \ , \ \ \  \cr
& i  [ \ H,\ {\cal S}^* ] =  - {\cal Q}^*, \ \ \ \ i [ \ P_-,  {\cal S} \ ]  = -Q_{0} ,
\ \ \ \ \ \  i \ [ D \ , \ {\cal S} \ ] \ = \ + {\cal S} \ , \ \ \ i [ \ J, \  {\cal S} \ ] \ =  -\frac{i}{2} {\cal S} \ , \cr
&  i  [ \ R, Q_0 \ ] = - i Q_0 , \ \ \ \ i  [\ R, Q_m ] = \frac{i}{3}Q_m , \ \ \ \ \  i [\ R \ , {\cal Q} \ ] \ = - i {\cal Q} \ , \ \ \  i  [ \ R \ , {\cal S} \ ] \ = - i {\cal S} \ .
\end{align}
Here, $R$ and $R^{mn}$ are $R$-symmetry generators, while ${\cal Q}$ and ${\cal S}$ are the dynamical supersymmetry and the superconformal generators, both of which are SU(2)$\times$SU(2) R-symmetry singlets.

We note that the non-relativistic superconformal algebra has a grading structure with respect to the dilatation operator $D$ and can be triangular-decomposed as
\begin{equation}
{\cal A}_+ \oplus {\cal A}_0 \oplus {\cal A}_-,
\end{equation}
where
\begin{eqnarray}
&& {\cal A}_+ = \{ \ P_- , \ {P}_+, \ H, \ {\cal Q},  \ {\cal Q}^*\ \} \nonumber \\
&& {\cal A}_0 = \{ \ J, \ R, \ R_{mn}, \ M, \ D, \ Q_0, \ Q_{0}^*, \ Q_m, \ Q_{m}^* \ \} \nonumber \\
&& {\cal A}_- = \{ \ G_-, \ {G}_+, \ K, \ {\cal S}, \ {\cal S}^*\ \}.
\end{eqnarray}
Related, for later purposes, we notice that the algebra has a non-trivial involution anti-automorphism of the algebra \cite{Nakayama:2008qm} given by
\begin{eqnarray}
&& w(J)= J, \qquad \ \ \ w(P_\pm)= {G}_\mp, \qquad \ w(G_\pm)={P}_\mp, \qquad  w(H) =-K, \nonumber \\
&& w(R) = R, \qquad \ w(D) = -D, \qquad \ w(M) = -M, \qquad \ w(K) = -H , \nonumber \\
&& w(Q_0) = iQ_{0}^*, \quad w(Q_{0}^*) = i Q_0, \qquad w(Q_m) = i Q_{m}^*, \ \quad \ w(Q_{m}^*)=iQ_m, \nonumber \\
&& w({\cal Q}) = i{\cal S}^*, \ \quad w({\cal Q}^*) = i {\cal S}, \ \qquad w({\cal S}) = i{\cal Q}^*,
\ \qquad w({\cal S}^*) = i{\cal Q}.
\end{eqnarray}
This anti-automorphism is much similar to the Belavin-Polyakov-Zamolodchikov (BPZ) conjugation of the relativistic two-dimensional conformal field theories and plays a central role in the representation theory of the non-relativistic superconformal algebra.

\section{Local Observables}

\subsection{operator-state correspondence}
We first recapitulate briefly the operator-state correspondence in the non-relativistic CFT \cite{Nishida:2007pj}. Consider a local operator $O(x)$, whose scaling dimension $\Delta$ and mass $\mu$ are given by the actions $i [D, O(x)] = -\Delta O(x)$ and $[M, O(x)] = \mu O(x)$, respectively. In the non-relativistic ABJM theory, local operators must form a representation of the super-Schr\"odinger algebra. If $O(x)$ commute with $G_\pm$ and $K$ (as well as ${\cal S}$ in the superconformal case), we call it the {\sl primary operator} \ :
\begin{eqnarray}
i[G_\pm,O(x)] = 0, \qquad i[K,O(x)] = 0, \qquad i \{ {\cal S}, O(x)] =0 \ .
\end{eqnarray}
A tower of {\sl descendant operators} is then obtainable by acting $i P_\pm := -\partial_\pm$ and $i H := \partial_t$ repeatedly on a given primary operator. Within each tower, the primary operator has the lowest scaling dimension. This is in accord with the grading structure with respect to $D$. In so far as $\mu$ is nonzero, we can always reach the primary operator by acting $G_\pm$ and $K$ on the descendants of each tower. Therefore, each tower is built upon a single primary operator, so we focus on the primary operators in analyzing the operator-state correspondence.

The operator-state correspondence in the non-relativistic CFT assigns to each primary operator an energy eigenstate of the system in a harmonic potential. In other words, we take the Hamiltonian to be
\begin{eqnarray}
 \mathcal{H} = H + K \ ,
\end{eqnarray}
where the special conformal generator $K$ provides a harmonic potential around the origin.
We see that the state $ |O\rangle = e^{-H}O(x=0)|0\rangle$ creates the energy eigenstate of $\mathcal{H}$
\begin{eqnarray}
\mathcal{H} |O\rangle = e^{-H} d_O O(x=0)|0\rangle = d_O |O\rangle \ ,
\end{eqnarray}
where $d_O$ is the scaling dimension of the primary operator:
$i[D,O(x)] = -(2t\partial_t + {\bf x} \cdot {\bf \partial} + d_O)O(x) $.
Note that the Hilbert space of this harmonic oscillator system has a natural involution anti-automorphism defined by the usual quantum-mechanical Hermitian conjugation of the harmonic oscillator. This anti-automorphism agrees well with the BPZ-like conjugation defined in section 2.3. Thus it is legitimate to use the unitary representation of the non-relativistic superconformal algebra based on the BPZ-like conjugation.

The argument here relies on the assumption that the operator $O$ does not annihilate the vacuum. Were $O$ annihilating the vacuum, we could instead consider $O^\dagger$ that would not annihilate the vacuum and study the corresponding state as long as $\mu$ is nonzero. Later, when we discuss operators with $\mu=0$, we shall encounter an important subtlety in extending this operator-state correspondence to the $\mu=0$ sector.

\subsection{zero-norm states and correlation functions}
One distinguishing feature of the non-relativistic ABJM theory we would like to show is that all the gauge invariant local observables (operators) have zero norm. In addition, all the correlation functions among them are trivial. This will eventually suggest a very peculiar nature of candidate gravity dual theory: the corresponding bulk fields should have zero norm as well. Nevertheless, we claim that these zero-norm states are non-trivial, as we will discuss further below.

There are several ways to see why all gauge invariant local operators must have zero norm in the non-relativistic ABJM theory.\footnote{We may consider a charge source such as monopole operator to avoid this constraint. The possibility will be discussed in section 5.} The crucial observation to this is that the total mass charge
\begin{eqnarray}
M = m\int\! \rmd^2 x \, \rho \ \qquad \mbox{where} \qquad
\rho = \mathrm{Tr}(\phi_A^\dagger \phi^A + \psi^{\dagger A} \psi_A) \ \label{masso}
\end{eqnarray}
measures matter contribution to the electric charge of the diagonal U(1) of the U$(N)\times$U($N$) gauge group.
As a consequence, if $O(x)$ is a {\sl gauge invariant} local (primary) operator, it must have zero electric charge and hence zero total mass: $i[M,O(x)]=0$.
This implies that, from the commutation relation $i[G_\mp,P_\pm] = 2M $ and the unitary representation of the superconformal algebra under the  BPZ-like conjugation $w(G_+) = P_-$ (see \cite{Nakayama:2008qm} for the details), which is compatible with the state-operator correspondence in the non-relativistic conformal field theory \cite{Nishida:2007pj}, $\vert\!\vert \ P_\pm|O \rangle \ \vert\!\vert^2 = 2\mu \vert\!\vert \ |O\rangle \ \vert\!\vert^2$ ought to vanish identically. Therefore, if $\mu$ of $O(x)$ is zero, $P_\pm |O\rangle$ has zero norm. If we also assume $P_\pm$ does not annihilate $O$, $|O \rangle$ must be a zero-norm state. Here, we have used the fact that, by definition, the primary states (created by the primary operators acting on the vacuum) are annihilated by $G_\pm$.

In this argument, we assumed that there is a unique Galilean vacuum, and $P_\pm|O\rangle$ states (or more precisely the corresponding operators $i[P_\pm,O(x)]$ ) are non-trivial. This assumption is reasonable since $i[P_\pm,O(x)] = \partial_\pm O(x)$ and the local operator $\partial_\pm O(x)$ is generically non-zero. In this way, we conclude that the state $|O\rangle$ itself must have zero norm.

An alternative way to see this is that any gauge invariant operator $O(x)$ is necessarily constructed out of {\it both} creation operators and annihilation operators. For example, $\mathrm{Tr}(\psi^{\dagger A}\phi^B)$ is constructed out of a creation operator $\psi^{\dagger A}$ and an annihilation operator $\phi^B$. Recall that the creation operators have positive total mass (positive diagonal U(1) electric charge) and the annihilation operators have negative total mass (negative diagonal U(1) electric charge), so we have to combine both fields to construct $\mu = 0$ gauge invariant operator. However, since the annihilation operator acting on the Galilean vacuum annihilates it ({\it i.e.} $\phi^B|0\rangle = 0$), the vacuum is annihilated by such `massless' gauge-invariant operators ({\it i.e.} $\mathrm{Tr}(\psi^{\dagger A}\phi^B) |0\rangle = 0$). This asserts that the corresponding states are zero-norm states.

In addition, we can show all the correlation functions among gauge invariant local operators are trivial. We will study the correlation functions such as
\begin{eqnarray}
\langle O_1(x_1) O_2(x_2) \cdots O_n(x_n) \rangle \ . \label{npint}
\end{eqnarray}
As we have seen, they are all zero-norm states: $\langle O_a(x)O_b(0) \rangle = 0$. The key point to show the triviality of \eqref{npint} is that, in the non-relativistic system, all the propagators of elementary fields in the perturbation theory are retarded ones. It is then easy to convince ourselves that, in perturbative evaluation of \eqref{npint}, we have to encounter backward propagators in time at least once, and such amplitudes vanish identically.\footnote{We can draw the same reasoning in the Euclidean path integral by the contour rotation argument. Physically, this corresponds to the statement that in the non-relativistic theory propagator consists only of retarded part and there is no particle anti-particle pair processes.} We would like to emphasize that the argument here does not rely on any supersymmetry: non-supersymmetric operators are not protected against radiative corrections in general, but still they remain zero-norm states. Possible renormalization of operators cannot possibly change the total mass charge.

\subsection{zero-norm states in gravity dual}
While there is yet no known gravity dual theory proposed for the non-relativistic ABJM theory, we could
understand some qualitative nature of these zero-norm states in the phenomenological non-relativistic gravity dual proposed in \cite{Son:2008ye}\cite{Balasubramanian:2008dm}. The metric of gravity dual is given by
\begin{align}
\rmd s^2 = -\frac{2(\rmd u)^2}{z^4} + \frac{-2\rmd u \rmd v + \rmd {\bf x}^2 + \rmd z^2}{z^2} \ ,
\end{align}
where the $v$-direction is compactified with radius $R$. The $n$-th Kaluza-Klein states have mass $\mu = n/R$. We shall take the limit $R \rightarrow 0$. In this limit, the only allowed state is the $n=0$ zero-mode. This fits to our assertion that there is no gauge invariant local operator with nonzero $\mu$ in the non-relativistic ABJM theory \footnote{When we consider monopole operators later, we shall see that the choice $1/R = k$ and $\mu = n k$ yields the correct spectrum.}.

Following \cite{Son:2008ye}, let us study a minimally coupled scalar field in the background
\begin{align}
S= - \int \rmd^5 x \sqrt{-g} (g^{\mu\nu} \partial_\mu \phi^* \partial_\nu \phi + m_0^2 \phi^* \phi) \ .
\end{align}
As said above, we are interested in the zero-norm sates, and take the Kaluza-Klein mode with $P_u = \mu = 0$. The action is reduced to
\begin{align}
S = \int \rmd^4 x \, \rmd z \, \frac{1}{z^{5}}(z^2\phi^* \partial^2_{\bf x} \phi -m_0^2 \phi^*\phi)\  . \label{raction}
\end{align}
Notice that there is no term with time derivative left.

The equation of motion is readily solved by Fourier transform as
$\phi_\pm (z, {\bf k}, \omega) = z^{2}K_{\pm \nu} (|{\bf k}|z)$ by the modified Bessel function with $\nu = \sqrt{m_0^2 + 4}$. By studying the asymptotics of $\phi_+$ and by using the GKPW prescription, we obtain the two-point correlation function of the operator $O(x)$ dual to $\phi$ in momentum space as
\begin{align}
\langle O({\bf k}, \omega) O(- {\bf k}, - \omega) \rangle \sim (|{\bf k}|^2)^{2\nu} \ .
\end{align}
Note that there is no dependence on the energy $\omega$. Due to this absence of $\omega$-dependence in the two-point function, going back to the coordinate-space gives $\delta(t)$ in the coordinate space. This is a manifestation of the fact that these are zero-norm states.

This elementary computation can be generalized to higher-point correlation functions. Since \eqref{raction} does not have any time derivative, the dynamics is frozen. As a result, any correlation functions among the local operators with $P_u = \mu = 0$ computed in the GKPW prescription contain string of delta-functions of $t$. This implies that the correlation functions are topological at best, as expected from the discussions based on Schr\"odinger algebra.

It should be remarked that the states with $P_u = \mu = 0$ is actually quite subtle in the framework of the discrete light-cone quantization. A prescription is that they are defined as the limit $P_u = \mu \to 0$, but then these states are lifted into ultraviolet infinitum of the light-cone Hamiltonian. Normally, they decouple from finite energy excitations but there are also known situations that such ultraviolet decoupling does not work. For example, suppose the theory under consideration has an anomalous global symmetry when coupled to gauge theory or gravity. Then, the global charge violation can be seen via pair production in the background of appropriate gauge or gravitational field background. In the conventional discrete light-cone quantization, where pair production / annihilation is not allowed, such a process seems not possible at all. It turns out that the states with $P_u = \mu \rightarrow 0$ are entirely responsible for reproducing the correct global charge anomaly. For details, we refer to \cite{Ji:1995ft}.

\subsection{supersymmetry protected operators}
In our considerations in this section so far, the supersymmetry did not play any role. We now classify local observables further in terms of the non-relativistic superconformal algebra. In particular, we are interested in local operators protected by the supersymmetry as well as the non-relativistic conformal symmetry. Unitary representations of the non-relativistic superconformal algebra (under the BPZ conjugation) was studied in \cite{Nakayama:2008qm}, where the unitarity bound as well as the null vector conditions were also presented
under the assumption $\mu \ne 0$ in the representation theory.  We shall utilize the results of \cite{Nakayama:2008qm}, exercising a little care because we are primarily focusing on `massless', zero norm states.

For instance, the mass or U(1) charge density $\rho$ in (\ref{masso}) is naturally protected. It also contains a null descendant at level-2 due to the conservation law $\partial_t {\rho} + \partial_\pm j^\mp = 0$, where $j_\pm$ is the momentum or U(1) current density. Such classes of null conditions were not considered explicitly in \cite{Nakayama:2008qm} as they are specific to the zero-norm operators.\footnote{Actually by carefully taking $\mu \to 0$ limit of the representation theory studied in  \cite{Nakayama:2008qm} , we can show that the null condition fixes the conformal dimension of $\rho$ to be $2$ and protected. It indicates that representation theory studied previously needs a careful treatment for this limit. However, in this paper, we will not discuss details of a systematic representation theory for $\mu = 0$.}

The supersymmetry protected operators, on the other hand, are not affected by the condition of $\mu=0$. We can therefore use the classification studied in \cite{Nakayama:2008qm}  (see also \cite{Lee:2009mm}). In the weak coupling limit ($k\to\infty$), we can construct these chiral primary operators out of chiral single `letters' as shown in table \ref{tab:2}. In particular, the chiral primary operators satisfy the condition
\begin{eqnarray}
\Delta = d_0 +j_0 - \frac{3}{2} r_0 = 0\ .
\end{eqnarray}
The chiral primary states satisfying this condition is annihilated by ${\cal Q}^*$. There are also anti-chiral primary operators out of conjugates of the single letters shown in table \ref{tab:2}. They satisfy the condition
\begin{eqnarray}
\overline{\Delta} = d_0 - j_0 + \frac{3}{2} r_0 = 0\ .
\end{eqnarray}
We can then form gauge-invariant (anti)chiral primary operators from multiple product of these letters and
contracting their gauge indices so that a gauge singlet is formed. They are further decomposed to irreducible symmetric representations of the SU(2)$\times$SU(2) R-symmetry.

\begin{table}[tb]
\begin{center}
\begin{tabular}{c|c|c|c|c|c}
 Letters        & U$(N) \times$U$(N)$& $J$ & $R$ & $d$ & $R-2J$ \\
 \hline
 $\phi^a$&     $ ({\bf N}, \overline{\bf N}) $         & $0$ & $2/3$ &$1$ & 2/3 \\
  $\phi^\dagger_{a'}$&     $ (\overline{\bf N}, {\bf N}) $         & $0$ & $2/3$ &$1$ & 2/3 \\
  $\psi^{\dagger a}$&     $ (\overline{\bf N} ,  {\bf N}) $         & $-1/2$ & $1/3$ & $1$ & 4/3 \\
$\psi_{a'}$&     $ ({\bf N}, \overline{\bf N}) $         & $-1/2$ & $1/3$ & $1$ & 4/3 \\
$P_+$&     $0 $         & $-1$ & $0$ & $1$ & 2 \\
 \end{tabular}
\end{center}
\caption{List of the letters contributing to the index (hence $\Delta = 0$) for the non-relativistic ABJM theory. $a,{a}' = 1,2$ are indices for SU(2)$ \times$ SU(2) R-symmetry.}
\label{tab:2}
\end{table}%

For example, we have the single-trace operators
\begin{eqnarray}
{\cal O}^{a_1 \cdots a_L}_{a'_1 \cdots a'_L} := \mbox{Tr} \Big[\phi^{(a_1} \phi^\dagger_{(a'_1} \phi^{a_2} \phi^\dagger_{a'_2} \cdots \phi^{a_L)} \phi^\dagger_{a'_L)}\Big] \qquad (L=1, 2, 3, \cdots)
\end{eqnarray}
of conformal dimension $2L$. They are classified by the symmetric spin-$L$ representation of SU(2)$\times$SU(2).

It is interesting to compare the supersymmetry protected operators in the non-relativistic ABJM theory with the supersymmetry protected operators in the parent relativistic ABJM theory. First, the flat directions of the non-relativistic ABJM theory does not descend from the flat directions of the relativistic ABJM theory. For the latter, we need to turn off the mass deformation, else the chiral ring structure is trivial and no nontrivial protected operators can be formed. Equivalently, the mass deformation in the relativistic ABJM theory lifts up the Coulomb branch completely. However, once taking the non-relativistic limit of the mass deformed relativistic ABJM theory, we see from the scalar potential (\ref{bosa}) of the non-relativistic ABJM theory that there are flat directions. Therefore, the flat directions in these two theories are separated by mass deformation and non-relativistic limit and hence are a priori unrelated.
Second, the R-symmetries of the two theories are different. In the relativistic ABJM theory, we have SU(4) R-symmetries, so the chiral operators can be classified by the symmetric representations of SU(4). Here, we only have SU(2)$\times $SU(2) symmetry. In particular $\phi^a$ and $\phi^{a'}$ should be treated quite differently: the former is a chiral letter but the latter is not (it is an anti-chiral letter).

It may be that that not all the gauge invariant combinations of the letters in table \ref{tab:2} are chiral primary operators for {\sl finite} $k$. As in the relativistic counterpart, the scalar potential (\ref{bosa}) in the action may impose nontrivial descendant equations as equations of motion. In non-relativistic supersymmetry, it is unclear how to formulate these descendant equations. However, in the computation of the superconformal index, the Bose-Fermi cancelation and the resultant invariance of the index guaranteed that the index does not change in the 't Hooft limit, so we can use the $k\to \infty$ result without having to solve the constraint from the potential terms.

\section{Non-local Observables and Correlation Functions}
So far, we considered local operators and their correlation functions. There are also various gauge-invariant nonlocal operators. Among them, we shall now consider the Wilson loop operator and study their correlation functions.
\subsection{Wilson loops}
With gauge fields present in the non-relativistic ABJM theory, nonlocal observables can be constructed. In this work, we shall focus on Wilson loops and `t Hooft loops. In gauge theories, these loop operators played an important role: they are order parameter for distinguishing the confinement and the Higgs phases. Often, they are also important mathematically. For example, in the {\sl pure} Chern-Simons theory where there is no dynamical degrees of freedom, the Wilson loops and the `t Hooft loops --- they are the same --- define knot and link topological invariants. Such topological nature ceases to be the case once {\sl relativistic matter} is coupled to the Chern-Simons gauge field. The relativistic ABJM theory falls into this class. What about the case of coupling {\sl non-relativistic matter} such as the non-relativistic ABJM theory? We claim that this case is exceptional in so far as these loop operators are concerned. In this section, we will show that the Wilson / `t Hooft loops in the non-relativistic ABJM theory are still {\sl topological} even though it contains nontrivial dynamical degrees of freedom.

By adapting the strategy for constructing those in the relativistic ABJM theory \cite{Rey:2008bh} (see also \cite{Drukker:2008zx}), we can construct  the Wilson loops in the non-relativistic ABJM theory. Again, as there are two gauge fields $A_m, \overline{A}_m$ associated with the two U($N$) gauge groups, two independent Wilson loops are present:
\begin{eqnarray}
&& W[C] := \frac{1}{N} \mathrm{Tr}P \exp\left[i\int_C \rmd \tau (\dot{x}^{\ m} A_m + M_{A}^{\ B} \phi^A \phi_B^\dagger + N_{A}^{\ B} \psi_B\psi^{\dagger A}) \right] \nonumber \\
&& \overline{W}[C] := \frac{1}{N} \mathrm{Tr}P \exp\left[i\int_C \rmd \tau (\dot{x}^{\ m} \overline{A}_m + {M}_{B}^{\ A} \phi_A^\dagger \phi^{ B} - {N}_{A}^{\ B} \psi^{\dagger A} \psi_{ B})\right] \ . \label{NRwilson}
\end{eqnarray}
Here, $M_A^{\ B}$ and $N_A^{\ B}$ are so-called velocity matrices of the contour $C$. Because the non-relativistic limit $c \rightarrow \infty$, they are non-zero only for time-like contour, $x^\pm = \mathrm{constant}$. Otherwise, $M_A^{\ B}$ and $N_A^{\ B}$ are no longer dimensionless parameters \footnote{Recall that $D(\phi^A) = D(\psi_A) =1$, $D(t)=-2$ and $D({\bf x})=-1$ in the non-relativistic system.}.

As in the relativistic counterpart, ${W}$ and $\overline{W}$ of the non-relativistic ABJM theory are related by the generalized parity transformation, which not only reverses the spatial orientation but also exchanges the two U($N$) gauge groups as well as two SU(2) R-symmetry groups. In the following, we thus focus on $W[C]$. As again in the relativistic counterpart, we readily see that one linear combination of (\ref{NRwilson}) is dual to Type IIA fundamental string. We refer to \cite{Rey:2008bh} for details of how these two possible classes of Wilson loops are compatible with the AdS/CFT correspondence

We now claim that the correlation functions among the Wilson loops \eqref{NRwilson} are {\sl topological} in that they depend only on knot and link topologies. In fact, they are identical to the correlation functions in the {\sl pure} Chern-Simons theory, where no propagating degrees of freedom is present. The argument is simple and does not rely on supersymmetry or conformal symmetry. First, consider the loops with $M_A^{\ B}=N_A^{\ B}=0$ but with arbitrary contour $C$. As we have stressed, there is no particle-antiparticle pair production and annihilation processes in non-relativistic system. As such, no Feynman diagram with internal matter loops contributes to the Wilson loop correlation functions. Thus, the only Feynman diagram contributing to the Wilson loop amplitudes come entirely from the same set of diagram contributing to the pure Chern-Simons theory. The correlation functions must agree to all orders in perturbation theory.

When we choose the contour $C$ timelike ($x^\pm = \mathrm{constant}$), we are able to introduce constant matrices $M_A^{\ B}$ and $N_A^{\ B}$ as well. As we will see, such choices are important for obtaining supersymmetric Wilson lines. In the Wilson loop expectation value, these constant matrix allow additional scalar (or fermion) exchanges. However, again there is no contribution to the correlation functions because the insertion $ M_{A}^{\ B} \phi^A \phi_B^\dagger + N_{A}^{\ B} \psi_{B}\psi^{\dagger B}$ has a net zero particle number, so one of the retarded propagators to close a loop makes the amplitudes vanish.

The reason why exchange of the non-relativistic ABJM matter fields does not contribute to the static potential can be easily understood from the relativistic ABJM theory and non-relativistic reduction thereafter. To take the non-relativistic limit, we effectively introduce infinite mass to the matter fields, keep particle components of them while suppressing anti-particle components. Then, the causal exchanges of massive matters would not contribute to the force between the static sources. Similarly, there is no internal matter loop in the amplitudes because either they are too heavy or there is no virtual pair processes available and the amplitudes are suppressed.

\subsection{Implications to gravity dual}
From the Wilson loop in the non-relativistic ABJM theory, we can also learn features and characteristic of putative gravity dual to the non-relativistic ABJM theory. Consider a unknot Wilson loop whose contour is a rectangle of width $L$ and $T$. It is well known that the Wilson loop expectation value is independent of the size and the shape of the contour. In the limit $T \rightarrow \infty$, this Wilson loop measures static potential between a pair of quark and anti-quark separated by a distance $R$. On general ground, by the non-relativistic conformal symmetry (Schr\"odinger symmetry), one expects that the potential scales as
\begin{eqnarray}
V(R) \quad = \quad - {C(\lambda) \over R^2} \ . \label{expected}
\end{eqnarray}
On the other hand, it is well-known that unknot Wilson loop expectation value is {\sl independent} of shape and hence on $T, R$. Therefore, $C(\lambda) = 0$ and the static potential is absent in the non-relativistic ABJM theory. We do not yet know what the gravity dual to the non-relativistic ABJM theory is. Nevertheless, we can still argue for peculiarity of the putative dual theory. Recall that, from the viewpoint of Wilson loop - string duality, the static potential (\ref{expected}) between two charges is directly related to the interaction energy between two strings emanating from the boundary. The interaction is mediated by the graviton, the dilaton and the Neveu-Schwarz-Neveu-Schwarz 2-form potential. As the two strings are antiparallel, the force mediated by each of these fields is attractive. Therefore, absence of the static potential in the gauge theory side implies that each of the three forces are zero!

Evidently, the above result suggests that the fundamental string in gravity dual is tensionless. This is because, as in the relativistic counterpart, the string tension ought to set a universal 'charge' for the Newtonian gravity, the dilaton and the 2-form forces. In section 6, we shall find another indication from the study of the dilatation operator in the non-relativistic ABJM theory and the spin chain picture that a string in the gravity dual background is tensionless.

Summing up, from considerations of quantum aspects, we see that putative gravity dual to the non-relativistic ABJM theory must be very different from the traditional AdS/CFT correspondence of the relativistic counterpart.

\subsection{Supersymmetric Wilson lines}
Much like the local observables, it is important to understand under what conditions the Wilson loops in the non-relativistic ABJM theory are supersymmetric. To this end, we start with the time-like Wilson loop operators in the defining representation with the ansatz:
\begin{eqnarray}
W[C] = \frac{1}{N} \mathrm{Tr}P \exp\left[i\int \rmd \tau (\dot{x}^{\ m} A_m + M_{A}^{\ B} \phi^A \phi_B^\dagger + N_{A}^{\ B} \psi_B\psi^{\dagger A}) \right] \ .
\end{eqnarray}
where the velocity matrices $M^{\ B}_A$ and $N_A^{\ B}$ are assumed constant. We now analyze conditions the above Wilson loop preserves part of the kinematical, dynamical and conformal supersymmetries.

Consider first the kinematical supersymmetry. The variation of the Wilson loop under the kinematical supersymmetry gives the following conditions on the spinor parameters $\xi, \hat{\xi}$:
\begin{align}
\frac{1}{4m} \hat{\xi}_{Ab} -M_{A}^{\ B} \hat{\xi}_{Bb} +N_{b}^{\ c} \hat{\xi}_{cA} - iN_{b}^{\ c'} \xi_{c' A} &= 0 \cr
\frac{1}{4m} \xi_{Ab'} + M_{A}^{\ B} \xi_{Bb'} - i N_{b'}^{ \ b}\hat{\xi}_{bA} - N_{b'}^{\ c'}\xi_{c' A} &= 0\cr
\frac{1}{4m}\xi^{aB} +M_{A}^{\ B} \xi^{A a} - N_{c}^{\ a} \xi^{cB} - iN_{c'}^{\ a} \hat{\xi}^{c' B} &= 0 \cr
\frac{1}{4m} \hat{\xi}^{a'B} - M_{A}^{\ B} \hat{\xi}^{A a'} -i N_c^{\ a'} \xi^{cB} + N_{c'}^{\ a'} \hat{\xi}^{c' B} &= 0  \ .
\end{align}
We decompose the velocity tensors into su(2)$\oplus$su(2) parts, and parametrize the matrix as $M_{a}^{\ b} =  \alpha \delta_{a}^{b}$ and $M_{a'}^{\ b'} = -\alpha \delta_{a'}^{b'}$ as well as $N_{a}^{\ b} =  \beta \delta_{a}^{b}$ and $N_{a'}^{\ b'} = -\beta \delta_{a'}^{b'}$. To preserve the component $Q_0$, they have to satisfy
\begin{eqnarray}
\frac{1}{4m} - \alpha -\beta = 0 \ .
\end{eqnarray}
Similarly, to preserve the components $Q_{m}$, they have to satisfy
\begin{eqnarray}
\frac{1}{4m} + \alpha - \beta = 0 \ .
\end{eqnarray}
In the former case, one can preserve two real kinematical supercharges, while in the latter case one, can preserve eight real kinematical supercharges. For the special case with $\alpha = 0$ and $\beta = \frac{1}{4m}$, one can preserve all the ten real kinematical supercharges.

Consider next the dynamical supersymmetry ${\cal Q}$. It is straightforward to see that one needs to set $N_{A}^{\ B} = 0$ to preserve the dynamical supersymmetry. The supersymmetry conditions now read
\begin{align}
{1 \over 4m} \zeta_{Ab} &= -M_{A}^{\ B} \zeta_{Bb} \cr
{1 \over 4m} \hat{\zeta}_{Ab'} &= + M_{A}^{\ B} \hat{\zeta}_{B b'} \cr
{1 \over 4m} \hat{\zeta}^{a B} &= + M_{A}^{\ B} \hat{\zeta}^{A a} \cr
{1 \over 4m} \zeta^{a'B} &= -M_{A}^{\ B} \hat{\zeta}^{A b'} \ \ .
\end{align}
We find that ${\cal Q}$ supersymmetry is preserved provided we choose $M_{a}^{\ b} = \delta_{a}^{b}$ and $M_{a'}^{\ b'} = -\delta_{a'}^{b'}$.

In summary, by choosing $M_{a}^{\ b} = -{1 \over 4m} \delta_{a}^{b}$ and $M_{a'}^{\ b'} = + {1 \over 4m} \delta_{a'}^{b'}$, we can preserve eight kinematical supersymmetries $Q_m$ as well as two dynamical supersymmetries ${\cal Q}$. For these choices, one easily checks that the conformal supersymmetry ${\cal S}$ is always broken.

We can investigate other supersymmetric Wilson loops which are not necessarily timelike. For example, it is obvious that the spacelike Wilson lines:
\begin{eqnarray}
W[C] = \frac{1}{N}\mathrm{Tr} P \exp\left(i\int \rmd \tau \ \dot{x}^{ \pm} A_\pm \right)
\end{eqnarray}
preserve all the ten kinematical supersymmetry simply because $A_\pm$ are invariant under the kinematical supersymmetry transformations. However, it is not possible to construct the Wilson lines that preserve the dynamical supersymmetry at the same time.

Finally, we consider the case with $\dot{x}^m=0$. The contour $C$ is a point in spacetime and gives rise to Wilson loop 'instanton' with
\begin{eqnarray}
W[C] = \frac{1}{N} \mathrm{Tr}P \exp \left(i\int \rmd \tau \  M^{\ B}_A \phi^A \phi^\dagger_B \right) \ .
\end{eqnarray}
These Wilson loops are actually a local operator in (2+1) dimensions.
There are several possibilities. For example, one can choose $M^{\ B}_A = \delta^{B}_1 \delta_A^1$ preserves four kinematical supersymmetries (combination of $\epsilon_3 + i\epsilon_4$ and $\epsilon_5 - i\epsilon_6$), or choose $M^{\ B}_A = \delta^{B}_1 \delta_A^{1'} $ preserves two kinematical supersymmetries (combination of $\epsilon_3 + i\epsilon_4$). In addition, one can preserve holomorphic or anti-holomorphic half of the dynamical supersymmetry ${\cal Q}$ by choosing $B=b$ and $A=a'$ or vice versa. We see that they generate the chiral (or anti-chiral) primary operators annihilated by ${\cal Q}^*$ or ${\cal Q}$, respectively. More generally, supersymmetric Wilson lines are generating functions of chiral primary operators. This is quite similar to the situation in the relativistic ABJM theory \cite{Rey:2008bh}.

\section{Nontrivial Observables and Correlation Functions}
This result of the previous sections is curious from the viewpoint of the non-relativistic AdS/CFT correspondence. Local observables in CFT are in one-to-one correspondence with bulk fields in AdS, and the correlation functions in CFT are computable from the bulk field amplitudes in AdS by adapting the GKPW prescription (see \cite{Son:2008ye}\cite{Balasubramanian:2008dm}\cite{Fuertes:2009ex}). What we showed so far is that all the bulk fields have zero norm and that fundamental string have topological interactions only. Does this mean that the non-relativistic ABJM theory itself is trivial or topological with little or no dynamical contents?

We claim, nevertheless, that the gauge invariant observables in the non-relativistic ABJM theory are nontrivial, and so should be the dual gravitational theory as well. After all, many of the zero-norm states are familiar quantities such as energy momentum tensor or conserved currents. While they create zero-norm states at the best, they are by no means trivial operators. Otherwise, we cannot construct charges of the non-relativistic conformal group from the outset. For example, the total density $\rho$ in \eqref{masso} has zero-norm, but certainly the mass operator $M$ is a nontrivial operator (and so should be $\rho$) \footnote{Related to this, in \cite{Nakayama:2008qm}, we have given a hypothetical finite norm to these ``massless states" to be counted in the index. Otherwise the index becomes trivial as well. They form a part of the protected operators, so this prescription makes sense physically.}.

Below, we illustrate that, along with Wilson loop operators, two classes of operators give rise to physically nontrivial correlations, evading the null property of operators with $\mu = 0$ we have shown in the previous section.

\subsection{monopole operators}
Recalling that crux of the argument for the null property of all conformal primary local operators lies in the observation that the particle number is nothing but the charge of the diagonal U(1) gauge group. As such, gauge invariant local operators contain both creation and annihilation operators of matter fields and annihilate the Galilean vacuum. Can there be any operator which can source the U(1) charges so that gauge
invariant operators involving only creation operators can be constructed? It is indeed possible in the
non-relativistic ABJM theory because the theory facilitates monopole operators and the Chern-Simons term.

Using the gauge fields in the non-relativistic ABJM theory, one can construct flux-changing, monopole operators. In fact, since the non-relativistic limit does not affect the gauge fields (except appropriate scaling of the gauge potentials), the monopole operators are the same for both non-relativistic and relativistic ABJM theory. In short, the monopole operator is a local operator in $(2+1)$ dimensions, defined as a specification of boundary condition for gauge fields: throughout $\mathbb{S}^2$ around the location of the operator, a quantized U$(N)$ gauge flux emanates. We specify the gauge flux in term of the singular magnetic monopoles \cite{Goddard:1976qe}. Canonically, they are labeled by magnetic charges of the Cartan subalgebra:
\begin{eqnarray}
F_{mn} = Q \epsilon_{mnp} {x^p \over 4 \pi |x|^3} \qquad \mbox{where} \qquad
Q = \mbox{diag}(q_1, \cdots, q_N) \ .
\end{eqnarray}
Any generic magnetic monopole configuration can be brought to this form by U$(N)$ conjugation.
The monopole charge $Q$, written in terms of Cartan generators, should obey the Dirac-Schwinger-Zwanziger quantization condition:
\begin{eqnarray}
\exp ( 2 \pi i Q) = \mathbb{I} \qquad \rightarrow \qquad q^a \in \mathbb{Z}.
\end{eqnarray}
In the non-relativistic ABJM theory, the Chern-Simons term lets these monopole operators also charged and transform in irreducible representations of the gauge group. The monopole with flux $q^a, (a=1, \cdots, N)$ transforms in the representation of U($N$) with the highest weight given by $(kq^1, kq^2, \cdots, kq^N)$,
where we ordered the U(1) charges as $q^1 \ge q^2 \ge \cdots \ge q^N$. The representation is labeled by Young tableaux whose $a$-th low has $kq^a$ many boxes. Since there are two U($N$) gauge groups, we have two sets of monopole operators as the basic building block. In particular, they carry respective U(1) charges $k \sum_a q^a$. We shall denote the corresponding monopole operators as $W (kq^1, \cdots, kq^N), \ \overline{W}(kq^1, \cdots, kq^N)$, respectively.
We see that they are local operators of {\sl nonzero} mass $ \mu >0$. For the relativistic ABJM theory, it was shown that these monopole operators are crucial to find precise agreement of the index between the conformal field theory side and the gravity dual side at non-perturbative level \cite{seokkim}.

We can construct local operators utilizing the above monopole operators \cite{Aharony:2008ug}. Recall that in the non-relativistic ABJM theory Chern-Simons term yields the Gauss law constraints for the U$(N)\times$U($N$) gauge fields:
\begin{eqnarray}
F_{12} - \frac{2\pi}{k}\left(
\phi^A\phi_A^\dag-\psi_A\psi^{\dag A}
\right) = 0~, \qquad
\overline{F}_{12}- \frac{2\pi}{k}\left(
\phi_A^\dag\phi^A+\psi^{\dag A}\psi_A
\right) = 0~ . \label{gausslaw}
\end{eqnarray}
Therefore, an operator of non-zero U(1) particle number can be made neutral by attaching a monopole operator of an appropriate representation. In the simplest case of U(1)$\times$U(1) gauge group, the monopole operator is given by $e^{k\sigma}$, where $\sigma$ is the pseudoscalar field dual to the Chern-Simons gauge field that couples to the total mass $M$. Because of the shift symmetry of $\sigma \to \sigma + \Lambda$ under the gauge transformation, one can construct a gauge invariant operator by attaching the monopole operator to an otherwise gauge non-invariant operator with non-zero $M$. More generally, for nonabelian U$(N)\times$U($N$) gauge group, it is possible to construct a gauge invariant composite operator by attaching the above GNO monopole operator to a product of nonabelian matter fields having non-zero mass $M$. For example, the so-called dibaryon operator
\begin{eqnarray}
{\cal B}_L := \phi^{(a_1} \phi^{a_2} \cdots \phi^{a_{kL})} W(L, 0, \cdots, 0, 0) \overline{W}(L, 0, \cdots, 0, 0) \qquad (L=1, 2, \cdots)
\end{eqnarray}
and its hermitian conjugate
\begin{eqnarray}
{\cal B}^\dagger_L := \phi^\dagger_{(a_1} \phi^\dagger_{a_2} \cdots \phi^\dagger_{a_{kL})} W(L, L, \cdots, L, 0) \overline{W}(L, L, \cdots, L, 0) \qquad (L=1, 2, \cdots)
\end{eqnarray}
are gauge invariant local operators carrying {\sl non-zero} mass $\mu = Lk$ and scaling dimension $\Delta = 2Lk$, where $L = 1, 2, \cdots$. There are also dibaryon operators in which string of $(\phi^b \phi^\dagger_{b'})^\ell, (\ell=1,2,\cdots)$ are inserted in front of various $\phi$'s in the operator. They form a continuum of the spectrum $\mu = Lk$ and $\Delta = 2Lk + 2\ell \ge 2Lk$. The highest weight states are supersymmetry protected and, for these states, the monopole operator does not contribute to scaling dimension.

Notice that in all the dibaryon operators the mass is different from the U(1) gauge charge: the U(1) charge of the matter-monopole composite operator is zero because contribution of the monopole operator have U(1) charge opposite to the matter fields. In fact, this is the distinguishing feature of local operators made out of monopole operators. The Gauss' law constraints are U(1) charge neutrality conditions for the sum of matter contribution and monopole contribution (induced by the Chern-Simons term). Therefore, they provide a class of operators that are gauge invariant yet have nonzero $\mu$ and non-vanishing norm. Below, because of this, we shall see that correlation functions involving these operators are physically nontrivial.

\subsection{bi-local and multi-local operators}
The second class of operators that can yield physically nontrivial correlations is the bi-local operators. They are defined by gauge invariant combinations words at two widely separated space-time locations and open holonomy lines connecting between them:
\begin{eqnarray}
W^A_{A'}(x, y) := \mathrm{Tr}\left[P \exp\left(i\int_x^y \rmd x^m \, {A}_m \right)\phi^A (x) P \exp\left(i\int_y^x \rmd x^m \, \overline{A}_m \right) \phi^\dagger_{A'} (y)\right] \ , \label{bilocal}
\end{eqnarray}
where $\phi, \phi^\dagger$ may be replaced by $\psi, \psi^\dagger$.
Here, for simplicity, we have omitted possible scalar (fermion) part in the Wilson lines by setting $M_A^{\ B}=N_{A}^{\ B}=0$. More generally, we can introduce the multi-local operators of the form
\begin{eqnarray}
W^{A_1 \cdots A_{2n-1}}_{A'_2 \cdots A'_{2n}} (x_1, \cdots, x_{2n}) &:=& \mathrm{Tr} \left[P \exp \left(i \int_{x_1}^{x_{2n}} \rmd x^m \, {A}_m \right) \phi^{A_1} (x_1) \ P \exp \left( i \int_{x_1}^{x_2} \rmd x^m \, \overline{A}_m \right) \phi^\dagger_{A'_2} (x_2)
\right. \cdots \nonumber \\
&& \hskip1cm \cdots \left. \phi^{A_{2n-1}}(x_{2n-1}) P \exp \left(i \int_{x_{2n-1}}^{x_{2n}} \rmd x^m \, \overline{A}_m \right) \phi^\dagger_{A'_{2n}} (x_{2n}) \right] \ ,  \label{multilocal}
\end{eqnarray}
where again $\phi, \phi^\dagger$ may be replaced by $\psi, \psi^\dagger$.

Nonlocal operators analogous to (\ref{bilocal}, \ref{multilocal}) can be constructed from the Wilson loop operators by first promoting the velocity tensors $M_A^B, N_A^B$ to contour-dependent function, taking functional derivative of the Wilson loop with respect to these velocity tensors, and finally setting $M_A^B, N_A^B$ to zero. It yields a class operators where a pair of words $\phi \phi^\dagger$ or $\psi \psi^\dagger$ is
distributed around the Wilson loop contour $C$.

The operators (\ref{bilocal}, \ref{multilocal}) are, however, more general than the Wilson loop operators in that the two primary words $\phi, \phi^\dagger$ are distributed around a (not necessarily simple) contour, connected by the open holonomy lines in gauge invariant manner. This marks a significant departure from the
Wilson loop operator since the pair of words $\phi \phi^\dagger$ or $\psi \psi^\dagger$ have vanishing U(1) charge and hence $\mu =0$, while the word $\phi$ and $\phi^\dagger$ or $\psi$ and $\psi^\dagger$ have non-vanishing U(1) charge and hence $\mu \ne 0$ locally. As for the dibaryon operators, we shall see that these multi-local operators yield physically nontrivial operators.

\subsection{mixed correlation functions}
We note that the distinguishing feature of the dibaryon and the multi-local operators is that they have non-vanishing particle number and hence $\mu \ne 0$ locally. We now argue that this is sufficient to lead to physically nontrivial correlation functions.

Consider, for instance, the dibaryon operator ${\cal B}_L$ and its hermitian conjugate, ${\cal B}_L^\dagger$ at two locations in spacetime. In the planar approximation, the two-point correlation function is given by
\begin{eqnarray}
\Big\langle {\cal B}_{L} (t_1, {\bf x}_1) {\cal B}^\dagger_{L'} (t_2, {\bf x}_2) \Big\rangle
\sim \delta_{L, L'} \ \theta(t_1 - t_2) \ (G (t_1- t_2, |{\bf x}_1 - {\bf x}_2))^{2Lk} \Delta(t_1 - t_2, {\bf x}_1 - {\bf x}_2)^L\ , \label{baryoncorrelation}
\end{eqnarray}
where
\begin{eqnarray}
G(t, {\bf x}) = {1 \over t} \exp ( - i m |{\bf x}|^2/2t)
\end{eqnarray}
is the Schr\"odinger propagator and $\Delta(t, {\bf x})$ is propagator of the monopole operator of defining representation in (2+1)-dimensional spacetime.
Notice that, in the weak coupling limit $k \rightarrow \infty$, both the mass $\mu = Lk$ and the scaling dimension $\Delta = 2Lk$ have a large gap of order $O(k)$ from the local operators of mass $\mu=0$ and the scaling dimension $\Delta = O(1)$. We see that the correlation decays very fast with spatial separation of the two dibaryon operators. We interpret this as the semiclassical counterpart of the light-cone squeezed along forward time direction (because $c \rightarrow \infty$) in non-relativistic system.

It is easy to find similar behavior for correlation functions involving the multi-local operators. For example, consider two-point correlation between the bi-local operators (\ref{bilocal}). By gauge fixing
$A_\mu = 0$ along the contour defining one of the two operators, we have
\begin{eqnarray}
&& \left\langle W^A_{A'}(t_1, {\bf x}_1, \tilde{t}_1, {\bf y}_1) \ W^B_{B'}(t_2, {\bf x}_2, \tilde{t}_2, {\bf y}_2) \right\rangle \nonumber \\
&& \hskip1.5cm \sim \delta^A_{B'} \delta^B_{A'} \ \theta(\tilde{t}_2 - t_1) \theta(\tilde{t}_1 - t_2) \ G(\tilde{t}_2 - t_1, |{\bf y}_2 - {\bf x}_1|) G(\tilde{t}_1 - t_2, |{\bf y}_1 - {\bf x}_2|).
\label{bilocalcorrelation}
\end{eqnarray}
Extending to multi-local operators, we see that the correlation is given by Wick theorem for various pairs of $\phi, \phi^\dagger$ subject to the causal boundary condition of the Schrodinger propagators.

If the bi-local operator $W(x,y)$ is expanded in powers of the separation $(x-y)$, it is given in an infinite series of spin-$s$, twist-2 local operators $O^{(s)}(x)$. From the consideration of section 3, we see that they are null operators with $\mu = 0$ and their correlation functions must all vanish. The fact that the correlation function (\ref{bilocalcorrelation}) is in general nonzero suggests that the operator product expansion and the infinite resummation of $O^{(s)}$ in spin $s$ do not commute. This marks another distinction of non-relativistic system from relativistic counterpart.

There are further non-vanishing correlation functions. One may insert arbitrary numbers of local operators with $\mu = 0$ or arbitrary numbers of Wilson loop operators to the correlation functions (\ref{baryoncorrelation}, \ref{bilocalcorrelation}). Evidently, they are physically nontrivial. The pattern that emerges out of all these correlation functions is this: to have physically nontrivial correlation functions, one needs to insert {\sl some} operators with nonzero $\mu$. The reason why the null property of
section 3 is evaded in this case is that U(1) charge of an operator is given by a sum of matter contribution $\rho$ and flux contribution as shown in (\ref{gausslaw}) while mass of an operator is given by $\mu = m \rho$.
Obviously, if the gauge flux of the operator under consideration is nonzero, the two quantum numbers are
different and the null condition no longer follows.

Physically speaking, while there is no vacuum polarization in the non-relativistic ABJM theory, there are non-trivial scattering amplitudes. The insertion of the dibaryon or the multi-local operators is a gauge invariant way to introduce isolated charged particles. Similarly, the insertion of the Wilson loop is a gauge invariant way to introduce external charged probe.

\section{Quantum Aspects}
Quantum dynamics of the non-relativistic ABJM theory entails also many interesting questions. In this section, we mention briefly two specific aspects and relegate further investigation and other issues to separate papers \cite{ours}.
\subsection{remarks on Chern-Simons level shift}
The non-relativistic ABJM theory is conformally invariant, so there is no infinite renormalization to the Chern-Simons coefficient $k$, whose inverse plays the role of perturbative expansion parameter. However, there can in principle be a {\sl finite} renormalization. In {\sl pure} Chern-Simons theory, Witten~\cite{witten} claimed that the Chern-Simons coupling $k$ is shifted by
\begin{eqnarray}
k \qquad \rightarrow \qquad k + {1 \over 2} C_2(G){\rm sign}(k),
\end{eqnarray}
where $C_2(G)$ is the Dynkin index (quadratic Casimir operator) for the adjoint representation of the gauge group $G$. For U($N$), we have $C_2(G) = N$. It is now known that the finite renormalization depends on regularization scheme. In so-called Yang-Mills scheme, the scheme in which Yang-Mills term is included to the pure Chern-Simons theory rendering the gauge fields topologically massive, the shift is nonzero: one-loop vacuum polarization diagrams of the gauge and the Faddeev-Popov ghost fields contribute additively. This is the regularization scheme implicit to Witten's claim. In dimensional reduction scheme, the scheme in which loop diagrams are first reduced to integrals involving Lorentz scalars only and then dimensionally continued, the shift is zero: one-loop vacuum polarization diagrams of the gauge and the Faddeev-Popov ghost fields cancel each other.

In the relativistic ABJM theory, which has ${\cal N}=6$ supersymmetry, the {\sl finite} renormalization is absent for both types of regularization schemes but for different reasons. In the Yang-Mills scheme, contribution of superpartners of the gauge field is also nonzero and, when summed together, cancels off the contribution of gauge and ghost fields. In the dimensional reduction scheme, contribution of superpartners cancels off by themselves. In fact, it is the ${\cal N}=2$ supersymmetry that is sufficient to ensure cancelation of the finite renormalization.

Now, in the non-relativistic ABJM theory, all the matter fields {\sl including} the ${\cal N}=2$ superpartner of the gauge field becomes non-relativistic. Being so, they cannot contribute to the vacuum polarization of the gauge field. Suppose one adopts the Yang-Mills regularization scheme. Then, absence of the vacuum polarization implies that the theory now receives a finite renormalization of the Chern-Simons coefficient, as in the pure Chern-Simons theory. Since $k$ is also present in front of matter part of the Lagrangian, by supersymmetry, this then implies that all non-relativistic matter fields receive finite renormalization of the wave functions by $\sqrt{(k+N)/k}$. To be consistent with gauge invariance, there ought to be no higher loop corrections beyond one loop. Suppose one instead adopts the dimensional reduction scheme. Then, since there was no finite renormalization of pure Chern-Simons theory and since the vacuum polarization is zero
for each and every non-relativistic matter fields, there is {\sl no} shift of the Chern-Simons coefficient.
We thus find a peculiar situation that a regularization scheme needs to be chosen in order to define the non-relativistic ABJM theory at quantum level.

We do not have a satisfactory resolution to this issue, but have a tentative preference to the Yang-Mills scheme. Our reasoning stems from the following physical standpoint. The conformal fixed point of the ABJM theory (both relativistic and non-relativistic) is defined as an infrared fixed point of renormalization group flow. The Lagrangian of the nonconformal theory along the renormalization group flow contains the Yang-Mills term with nonvanishing coefficient off the fixed point. We should emphasize again that the limit of approaching the infrared fixed point is smooth in the relativistic ABJM theory but discontinuous in the non-relativistic counterpart.

Under the above caveat on using the Yang-Mills scheme, the finite renormalization $k \rightarrow (k+N)$ bears an interesting implication in search for gravity dual to the non-relativistic ABJM theory. In the relativistic ABJM theory, such a finite renormalization was absent, and the `t Hooft coupling $\lambda := \sqrt{N/k}$ can take an arbitrarily large value, rendering geometry of the gravity dual macroscopically large in string unit and so weakly curved. In the non-relativistic ABJM theory, we see that this is not possible. The `t Hooft coupling is now given by $\lambda := \sqrt{N / (N+k)}$, and can take ${\cal O}(1)$ at most. Therefore, geometry of the gravity dual ought to have string scale size and becomes strongly curved.
We see that, along with topological nature of interactions, the gravity dual features string-scale geometry.

\subsection{non-relativistic spin chain and dual string theory}
Analogous to the relativistic counterpart \cite{minahanzarembo, ours1}, \cite{ours2, ours3}, we can also study conformal dimensions of gauge invariant, single trace operators. We choose the ground state to a supersymmetry protected operator, for example,
\begin{eqnarray}
\vert 2L \rangle \qquad \leftrightarrow \qquad \mbox{Tr} \Big[\phi^1 \phi^\dagger_{1'} \phi^1 \phi^\dagger_{1'} \cdots \phi^1 \phi^\dagger_{1'} \Big].
\end{eqnarray}
Excited states above this ground state with higher conformal dimension are constructible simply by replacing $\phi^1$ to $\phi^2$ and $\phi^\dagger_{1'}$ to $\phi^\dagger_{2'}$. We have essentially two SU(2) spin chains alternatingly interweaved. As in the relativistic ABJM theory, these states correspond to magnon excitations in the alternating spin chain. We would like to find dilatation operator and its eigenvalues on these states at weak `t Hooft coupling limit. We expect these operators are in one-to-one correspondence with excited states of a single Type IIA string in yet-to-be-found gravity dual background.

At large $N$, only planar diagrams contribute. By inspection of the scalar potential (\ref{bosa}), it is evident that interactions among the matter fields do not contribute at all, simply because all planar diagrams involve pair creation/annihilation processes and they are prohibited in a non-relativistic system as this theory is. Interactions involving gauge fields are possible but they do not flip the SU(2) flavor 'spins' and only contribute to the ground-state energy. When summed up, these contribution ought to vanish since the dynamical supersymmetry protects the ground-state from acquiring nonzero energy. This means that all single trace operators, supersymmetry protected or not, have vanishing anomalous dimension to all order in perturbation theory. We conclude that planar contribution to the dilatation operator is simply an identity operator to {\sl all} orders in the `t Hooft coupling:
\begin{eqnarray}
H_{\rm planar} = \sum_{a=1}^{2L} \mathbb{I} + \cdots
\end{eqnarray}
and counts {\sl classical} scaling dimension for all single-trace gauge invariant states. Here, the ellipses are subleading contributions in the planar limit. At order $O(1/N)$, there are joining and splitting interactions changing the number of traces of the spin chain by one. At order $O(1/N^2)$, in addition to second-order trace-changing processes, there is also an interaction preserving the number of traces of the spin chain.

By interpreting the set of single-trace states as the excitation states of a single Type IIA string in gravity dual, we learn that the worldsheet dynamics is trivial. As the spin-chain Hamiltonian (as described by the dilatation operator) has no term exchanging letters (spins), there cannot possibly be a `magnon' excitation in the spin chain picture. This in turn implies that there cannot possibly be a propagating worldsheet excitation in the string picture. Roughly speaking, string is made of bits, but the bits do not interact each other for {\sl any} value of the string tension. Again, they exhibit topological character.

Once we take account of nonplanar contributions mentioned above, the spin chain ceases to be topological. This is evident from the Feynman diagrammatics. At one loop and beyond, there are nontrivial joining / splitting and exchange terms contributing to the dilatation operator. In this regard, the joining and splitting of the spin chain and the string interaction become nontrivial. The issue is outside the main focus of this paper, so we shall relegate further elaboration elsewhere \cite{ours}.

Finally, we remark that exactly the same argument applies also to the dibaryon operators, though this was the feature already present in the relativistic ABJM theory.

\section{Conclusion}
In this paper, we investigated gauge-invariant observables and their correlation functions in the non-relativistic ABJM theory. We showed that the local gauge invariant observables are all zero-norm states, and the correlation functions among them are trivial. We also showed that Wilson loops are topological in the sense that the matter fields impart no radiative corrections and their correlation functions coincide with those of the pure Chern-Simons theory, depending only on link and knot of the contour. We further showed that the theory is nevertheless non-topological since there are physically nontrivial correlation functions. We illustrated a few of them by utilizing monopole operators and multi-local operators.

Although it has not been demonstrated yet, it must be that the non-relativistic ABJM theory has a sensible gravity dual description. Indeed, massive deformation of the relativistic M2-brane gauge theory is known to have a dual geometry \cite{Bena:2004jw}. It remains to take gravity counterpart of the non-relativistic limit \footnote{Instead of this systematics, \cite{ooguri} attempts a bottom-up approach to construct an M-theory solution. The claimed solution, however, turns out not to have the requisite kinematical supersymmetry.}. The results of our paper suggests that this limit should lead to a very peculiar gravity dual. First of all, all the bulk fields must have zero norm with $\mu=0$. Second, the Wilson loops, most likely represented by the fundamental strings stretched toward the boundary, must be topological and tensionless. Third, the gravity dual have a geometry of string scale and string theory defined on it is topological. Fourth, some correlation functions involving monopole or multi-local operators must be physically nontrivial. Fifth, it should be noted that the non-relativistic conformal symmetry (Schr\"odinger symmetry) is not a part of the relativistic counterpart OSp$(6 \vert 4)$. Related to this, the moduli space of the non-relativistic ABJM theory is not related not in any direct way to the moduli space of the relativistic ABJM theory --- in fact, the latter was lifted completely as soon as the mass deformation is introduced.

We emphasize that these peculiar aspects may not be particular to the non-relativistic ABJM theory, and they may form a universality class to all non-relativistic conformal field theories. For instance, suppose a dual conformal field theory is given by a generic Chern-Simons-Matter theory where the total mass charge is a part of the gauge symmetry. Then the triviality of the correlation functions among local observables and the topological nature of the Wilson loops follow directly from the same argument as given in this paper. Notice that, in this argument, supersymmetry plays a minor role. Consequently, it seems very important to study other non-local objects, not conventionally studied in the AdS/CFT correspondence, to reveal the dynamics of such non-relativistic CFTs.

Finally, we would like to comment on embedding of the non-relativistic AdS/CFT backgrounds in the string theory proposed in \cite{Herzog:2008wg}\cite{Maldacena:2008wh}\cite{Adams:2008wt} and in the M-theory proposed in \cite{ooguri}. A common thread of their backgrounds is that they all contain a null conical singularity. One can then wrap any numbers of strings or M2 branes along the light-like compact directions, rendering them effectively tensionless. In this sense, their background is unstable and pathological as a ground state. A partial topological nature of the non-relativistic ABJM theory discussed in this paper might be related to the instability of their string background. Their singularity was removed once they introduced the finite temperature and finite density. It is easy to see that the topological feature of the non-relativistic ABJM theory no longer persists once we introduce finite temperature and finite density.

We strongly believe that this research stimulates further studies on the non-relativistic AdS/CFT and the (non)local observables therein and the search for string or M theory defined on putative holographic dual with fascinating topological structures.

\section*{Acknowledgements}

It is our pleasure to acknowledge stimulating discussions and correspondences with Dongsu Bak, Nikolay Bobev, David J. Gross, Seok Kim, Kimyeong Lee, Seongjay Lee, Hyunsoo Min, Hirosi Ooguri, Chang-Soon Park, Makoto Sakaguchi and Kentaroh Yoshida. SJR thanks warm hospitality of the Kavli Institute for Theoretical Physics and the Berkeley Center for Theoretical Physics during this work. The work of YN was supported in part by the National Science Foundation under Grant No.\ PHY05-55662 and the UC Berkeley Center for Theoretical Physics. The work of SJR was supported in part by SRC-CQUeST-R11-2005-021, the Korea Research Foundation Grant KRF-2005-084-C00003, the Korea Science and Engineering Foundation Grant KOSEF-2009-008-0372 and the U.S. National Science Foundation under Grant No. PHY05-51164 at KITP.




\end{document}